\documentclass[aps,prl,twocolumn,groupedaddress]{revtex4}

\usepackage{graphicx} 
\usepackage{dcolumn}% Align table columns on decimal point
\usepackage{bm}% bold math

\begin{document}
\title{Precise determination of the superconducting gap along the diagonal 
direction of Bi$_{2}$Sr$_{2}$CaCu$_{2}$O$_{8+y}$:   Evidence for 
extended $s$-wave gap symmetry }

\author{Guo-meng 
Zhao$^{*}$} 

\affiliation{ Department of Physics and Astronomy, California State 
University at Los Angeles, Los Angeles, CA 90032, USA}

\begin{abstract}
We report high-resolution spectra of the second 
derivative $-$d$^{2}$(Re$\Sigma$)/d$\omega^{2}$  of the real part of electron self-energy $\Sigma$ along the diagonal  direction (where  $\omega$ = $E_{F}-E - 
\Delta_{D}$ and $\Delta_{D}$ is the diagonal superconducting gap) for a nonsuperconducting La$_{1.97}$Sr$_{0.03}$CuO$_{4}$ (LSCO) crystal and a superconducting 
    Bi$_{2}$Sr$_{2}$CaCu$_{2}$O$_{8+y}$ (BSCCO) crystal with $T_{c}$ = 91 K.  The    
$-$d$^{2}$(Re$\Sigma$)/d$\omega^{2}$ spectrum of the nonsuperconducting LSCO
shows clear peak
features, which match precisely with those in the phonon
density of states obtained from neutron scattering.  Similarly, if we 
assign $\Delta_{D}$ = 7$\pm$1 meV for the
superconducting BSCCO, the
peak features in $-$d$^{2}$(Re$\Sigma$)/d$\omega^{2}$ of this compound
also match precisely with those in the phonon density of states and 
the tunneling spectrum. The 
present results rule out seemingly well accepted $d$-wave gap symmetry and strongly 
support an extended $s$-wave gap symmetry with eight line nodes.

\end{abstract}
\maketitle 
 
 The phenomenon of superconductivity involves the pairing of 
electrons into Cooper pairs \cite{BCS}.  The internal wavefunction 
(gap function) of these Cooper pairs obeys a certain symmetry which 
reflects the underlying pairing mechanism.  It is known that 
conventional superconductors (e.g., Pb and Nb) possess an $s$-wave gap 
symmetry that reflects the phonon mediated electron pairing 
\cite{BCS}.  On the other hand, the symmetry of the gap 
function in high-temperature superconductors has been a topic 
of intense debate for over fifteen years. 
Phase-sensitive experiments based on planar Josephson tunneling 
\cite{Review} appear to provide compelling evidence for $d$-wave 
order parameter (OP) symmetry.  Since the gap symmetry is the same as 
the OP symmetry in BCS-like superconductors, these experiments lead 
to a wide-spread belief that $d$-wave gap symmetry has been firmly
established for high-temperature superconductors. 
However, it is worth noting that these phase-sensitive 
experiments are probing the OP symmetry on surfaces/interfaces, 
which are found to be significantly underdoped \cite{Bet,Mann}.  The evidence 
for $d$-wave OP symmetry on 
the underdoped surfaces/interfaces is consistent with $d$-wave symmetry of the 
Bose-Einstein condensate of intersite oxygen-hole pairs 
\cite{Alex}, which are the primary charge carriers in underdoped 
cuprates \cite{Muller}.  Therefore, these surface- and phase-sensitive 
experiments do not provide conclusive evidence for $d$-wave gap 
symmetry in the bulk of high-temperature superconductors. 

On the other hand, several independent bulk 
sensitive experiments (e.g., penetration depth and thermal conductivity
measurements) have pointed to the existence of line 
nodes in the gap function \cite{Hardy,Jacobs,Lee,Chiao}.  
Qualitatively, these experiments are consistent with $d$-wave gap 
function with four line nodes.  Nevertheless, these data are also 
consistent with an extended $s$-wave gap function ($s + g$ wave), 
which has eight line nodes when the $g$-wave component is larger than 
the $s$-wave one.  Therefore, either $d$-wave or extended 
$s$-wave gap symmetry is supported by these bulk-sensitive 
experiments. 

In order to make a clear 
distinction between the $d$-wave and extended 
$s$-wave gap symmetries, it is essential to determine precisely  the 
superconducting gap along the diagonal 
direction (45$^{\circ}$ from the Cu-O bonding direction).  If the 
diagonal gap is substantial, the $d$-wave gap symmetry can be ruled 
out, and only the extended $s$-wave gap symmetry  is plausible. 
Several angle resolved photoemission spectroscopy (ARPES) experiments 
attempting to determine the magnitude of the diagonal gap ($\Delta_{D}$)
have led to contradictory conclusions.  From the midpoint shift of 
the leading edges 
of the normal and superconducting ARPES data obtained with 
a 30 meV energy resolution, Shen {\em et al.} \cite{Shen} 
found $\Delta_{D}$ $\simeq$ 0 for an overdoped 
Bi$_{2}$Sr$_{2}$CaCu$_{2}$O$_{8+y}$ (BSCCO)  with $T_{c}$ = 78 K 
and $\Delta_{D}$ = 2-12 meV for a nearly optimally 
doped BSCCO with $T_{c}$ = 86 K. By using the same 
criterion they estimated \cite{Shen} the anti-nodal gap $\Delta_{M}$ to be 
about 12 meV, 
which is a factor of 2.8 smaller than the maximum gap (34 meV) deduced 
from scanning tunnelling spectroscopy for an overdoped BSCCO 
with $T_{c}$ = 74.3 K (Ref.~\cite{Renner}). 
Using a fitting procedure for the ARPES spectra 
obtained with a 19 meV energy 
resolution, Ding {\em et al.}  showed $\Delta_{D} = 3.5\pm$2.5 meV 
(Ref.~\cite{Ding}) or $\Delta_{D} = 0\pm$3 meV (Ref.~\cite{Ding1}) in 
a BSCCO with $T_{c}$ = 87 K. With a better energy 
resolution (about 10 meV), Vobornik {\em et al.} found $\Delta_{D} = 
9\pm$2 meV for a heavily overdoped BSCCO with $T_{c}$ = 60 K 
(Ref.~\cite{Vob}).  On the other hand, from the midpoint shift of 
the leading edges of the spectra obtained with a 13-15 meV energy 
resolution,   Gatt {\em et al.} found $\Delta_{D} $ = 0  in an 
overdoped BSCCO with $T_{c}$ = 65 K (Ref.~\cite{Gatt}). 
These contradictory conclusions about 
the magnitude of  $\Delta_{D}$ may arise from a limited energy 
resolution and subjective criteria for extracting the gap. 
Therefore, the $d$-wave gap symmetry has not
been firmly established for high-temperature superconductors.

Here we report high-resolution spectra of the second derivative $-$d$^{2}$(Re$\Sigma$)/d$\omega^{2}$  of the real part of 
electron self-energy $\Sigma$ along the diagonal ($\Gamma-Y$) direction (where  $\omega$ = $E_{F}-E - 
\Delta_{D}$) for a nonsuperconducting La$_{1.97}$Sr$_{0.03}$CuO$_{4}$ 
(LSCO) crystal and a superconducting 
    Bi$_{2}$Sr$_{2}$CaCu$_{2}$O$_{8+y}$ crystal with $T_{c}$ = 91 K.  The  $-$d$^{2}$(Re$\Sigma$)/d$\omega^{2}$
    spectrum of  the nonsuperconducting LSCO
shows clear peak
features, which match precisely with those in the phonon
density of states obtained from neutron scattering.  Similarly, if we 
assign $\Delta_{D}$ = 7$\pm$1 meV for the
superconducting BSCCO, the
peak features in $-$d$^{2}$(Re$\Sigma$)/d$\omega^{2}$ of this compound
also match precisely with those in the phonon density of states and 
the tunneling spectrum, and with the peak features in the normal-state $-$d$^{2}$(Re$\Sigma$)/d$\omega^{2}$
spectra of several underdoped La$_{2-x}$Sr$_{x}$CuO$_{4}$ crystals. The 
present results rule out seemingly well accepted $d$-wave gap symmetry and strongly 
support an extended $s$-wave gap symmetry with eight line nodes.

\begin{figure}[htb]
    \includegraphics[height=6.2cm]{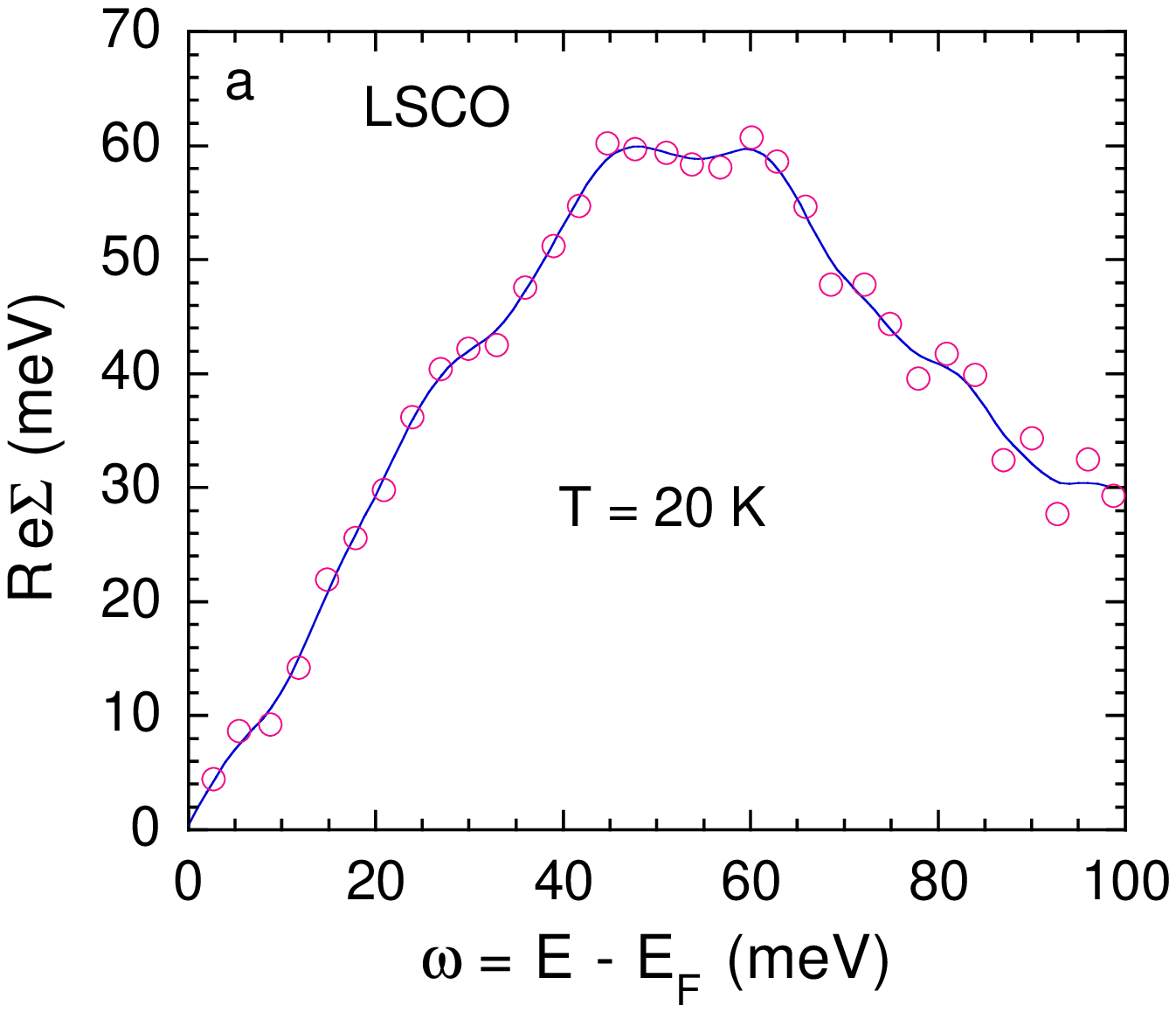}
	 \includegraphics[height=6.2cm]{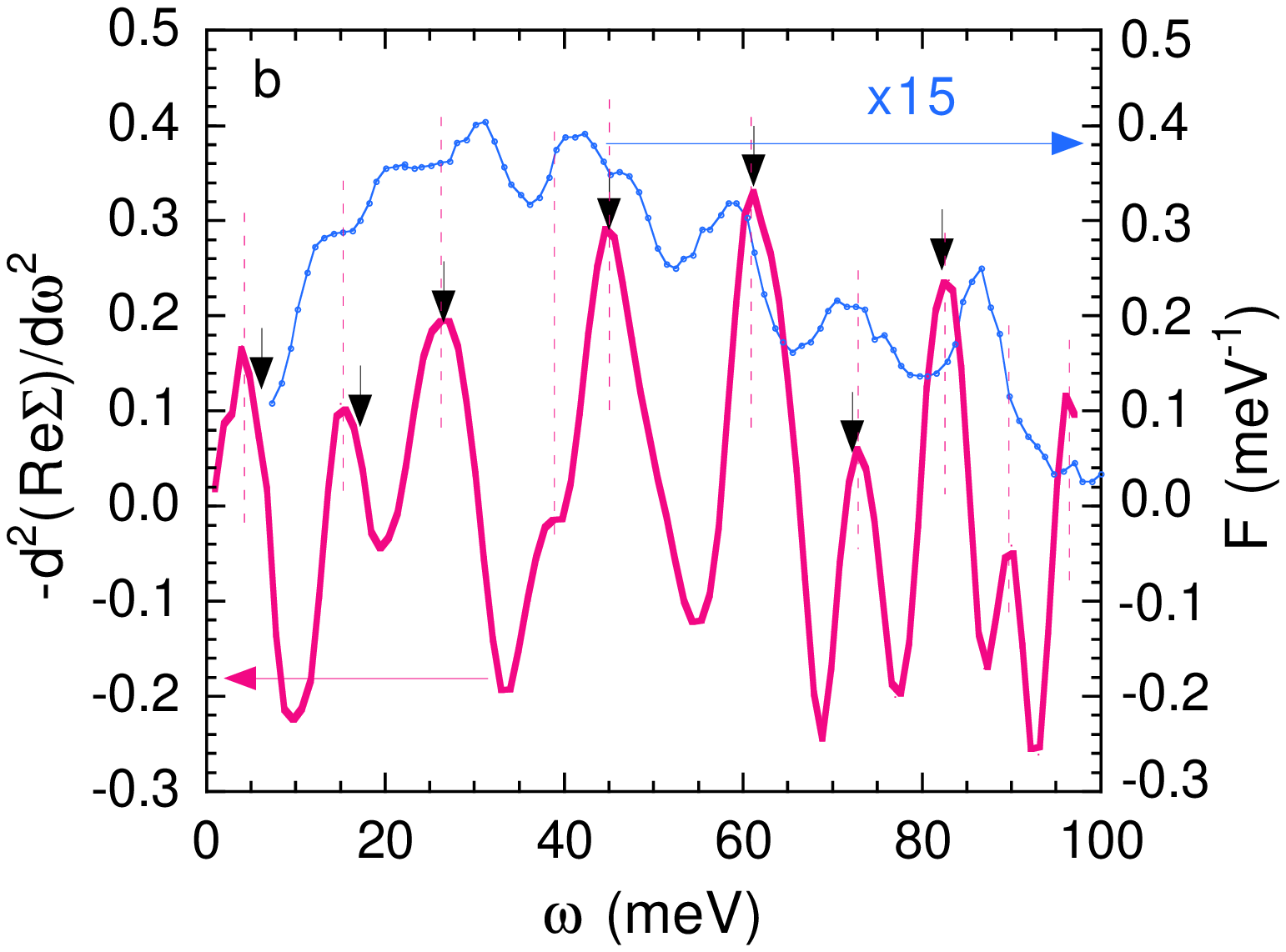}
	\caption[~]{a) The real part of the electron self-energy at 20 K for 
a nonsuperconducting La$_{1.97}$Sr$_{0.03}$CuO$_{4}$ crystal. The data 
are reproduced from 
Fig.~2 of Ref.~\cite{Zhou}. The solid line is a 
smoothed curve with 
a smoothing parameter $\Lambda$ = $-$2.5 (see text). b) 
$-$d$^{2}$(Re$\Sigma$)/d$\omega^{2}$  
spectrum (left scale)
obtained from the smoothed curve in a) and the phonon density of 
states (right scale) of La$_{1.85}$Sr$_{0.15}$CuO$_{4}$ at 20 K 
(Ref.~\cite{AraiLSCO}).  Downward arrows 
mark the 
positions of the boson modes
in the electron-boson spectral function obtained using a maximum 
entropy method \cite{Zhou} and vertical dashed lines  indicate the positions 
of the peaks in $-$d$^{2}$(Re$\Sigma$)/d$\omega^{2}$. }
\end{figure}

In Fig.~1a we plot the real part of the electron self-energy at 20 K for a 
nonsuperconducting 
La$_{1.97}$Sr$_{0.03}$CuO$_{4}$ crystal. The data are reproduced from 
Fig.~2 of Ref.~\cite{Zhou}. The electron self-energy  data are 
obtained from an ARPES spectrum, which is taken with an energy 
resolution of 
about 20 meV (Ref.~\cite{Zhou}). It is apparent that the electron 
self-energy shows fine 
structures associated with the bosonic modes coupled to electrons. In 
order to precisely determine the energies of the bosons, it is 
essential to take the second derivative of Re$\Sigma$. Before taking 
the second derivative of Re$\Sigma$, we need to obtain a smoothed 
curve for the data. We use a cubic spline interpolation method to 
smooth the data, which ensures no discontinuity in the first and 
second derivative. The smoothness of the interpolation is determined 
by a parameter $\Lambda$. The larger the  $\Lambda$ is, the smoother 
the curve is. The solid line in Fig.~1a is a smoothed curve with 
$\Lambda$ = $-$2.5. 

Fig.~1b shows the 
$-$d$^{2}$(Re$\Sigma$)/d$\omega^{2}$  
spectrum 
obtained from the smoothed curve in Fig.~1a. There are 10 peak 
features at 4.2 meV, 15.2 meV, 26.2 meV, 38.8 meV, 45.0 meV, 60.8 
meV, 72.8 meV, 82.5 meV, 89.6 meV, and 96.1 meV. The energies of the 
peak 
features match very well with the boson energies (marked by downward 
arrows) in the electron-boson 
spectral function obtained using a maximum entropy method 
\cite{Zhou}. Such excellent agreement indicates that the smoothed curve 
with $\Lambda$ = $-$2.5 represents a realistic average of the data. 

In order to show that these fine structures in  $-$d$^{2}$(Re$\Sigma$)/d$\omega^{2}$ are 
associated with strong electron-phonon coupling, we 
compare these fine structures with those in the phonon 
density of states. It is apparent that nearly all the fine structures match precisely with those in 
the phonon density of states. The double peaks at 82.5 meV and 89.6 
meV in $-$d$^{2}$(Re$\Sigma$)/d$\omega^{2}$  should match with the single 
broad peak at 86.0 meV in the phonon density of states. Since the low-energy fine strcutures in
self-energy 
are more difficult to resolve than the high-energy ones because of the
same thermal broadening (4.4$k_{B}T$) \cite{Dev}, the single
broad peaks at 26.2 meV and 45.0 meV in $-$d$^{2}$(Re$\Sigma$)/d$\omega^{2}$
should match with the double peaks in the phonon density of states.
The peak at 
about 4.2 meV is consistent with 
the specific heat data \cite{Art} of 
La$_{1.85}$Sr$_{0.15}$CuO$_{4}$, which reveal an Einstein mode at about 
4.3 meV.  Therefore, nearly all the peak features in 
$-$d$^{2}$(Re$\Sigma$)/d$\omega^{2}$ precisely match with those in 
the phonon density of states, which clearly indicates that the fine 
structures in $-$d$^{2}$(Re$\Sigma$)/d$\omega^{2}$ are indeed caused 
by strong electron-phonon coupling. 

\begin{figure}[htb]
    \includegraphics[height=6.2cm]{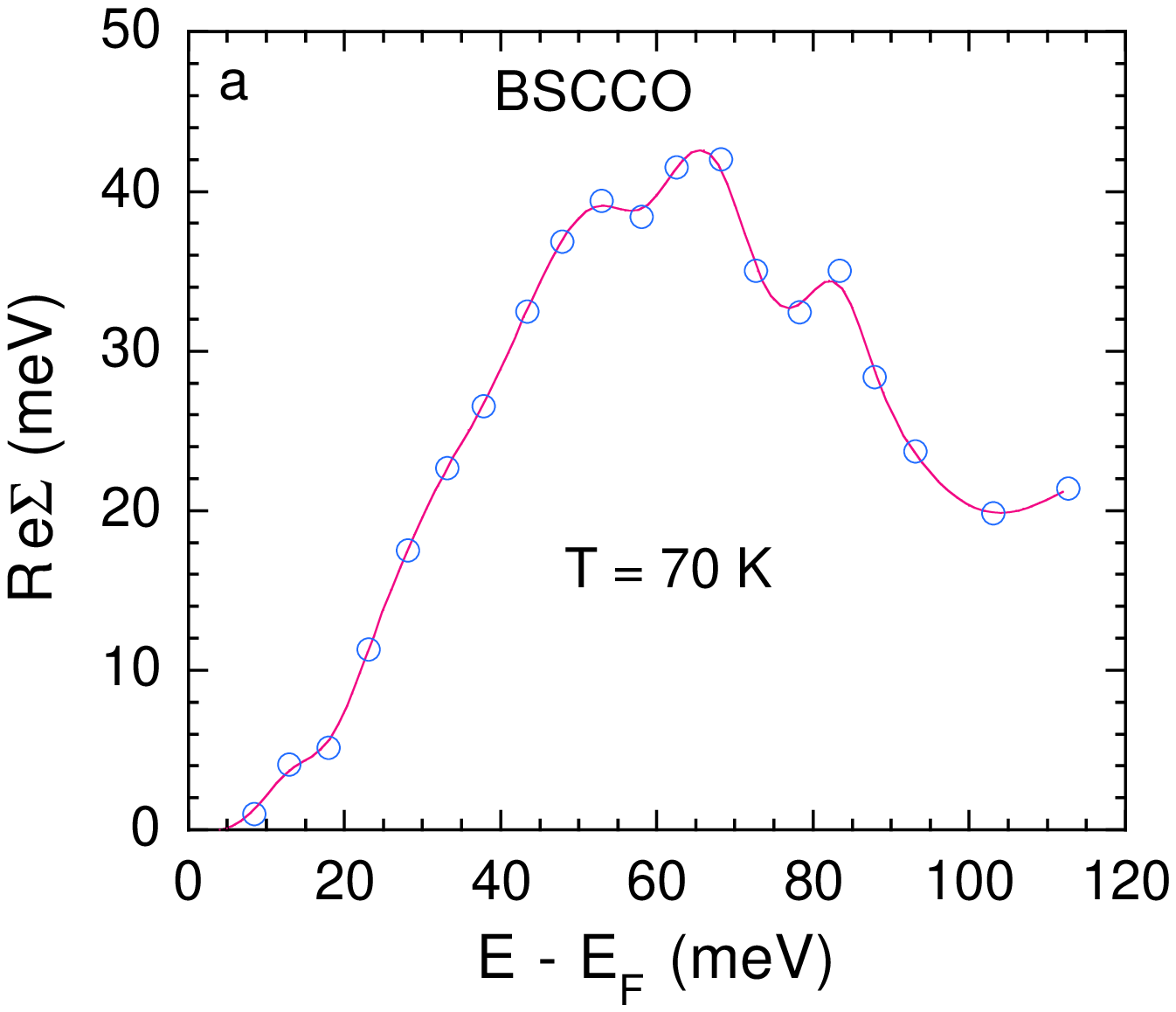}
	 \includegraphics[height=6.2cm]{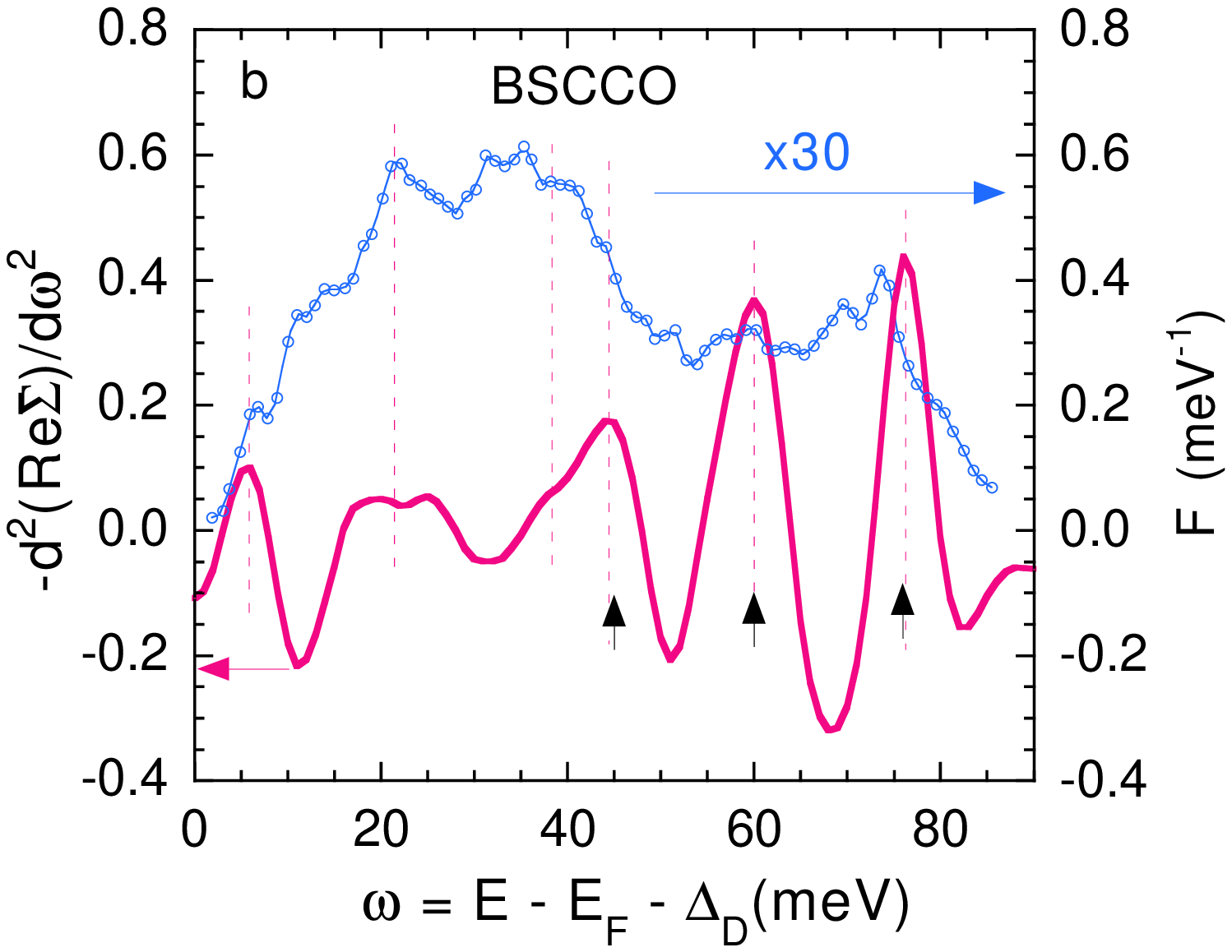}
	\caption[~]{a) The real part of the electron self-energy at 70 K for 
a superconducting BSCCO crystal with $T_{c}$ = 91 K.  The data are 
reproduced from Ref.~\cite{Johnson}.  The solid line is a smooth curve obtained 
using the cubic spline interpolation with $\Lambda$ = -4.  b) The 
$-$d$^{2}$(Re$\Sigma$)/d$\omega^{2}$ spectrum of a superconducting BSCCO crystal (left scale) and the phonon density of 
states (right scale) of BSCCO (Ref.~\cite{Renker}).  Here we have
assiged $\omega$ = $E_{F}-E - 
\Delta_{D}$ with $\Delta_{D}$ = 7 meV. Upward arrows 
mark the 
positions of the peaks
in the normal-state $-$d$^{2}$(Re$\Sigma$)/d$\omega^{2}$ spectra in
underdoped superconducting LSCO crystals \cite{Zhou} and vertical dashed lines  indicate the positions 
of the peaks in the $-$d$^{2}$(Re$\Sigma$)/d$\omega^{2}$ spectrum of
BSCCO. }
\end{figure}

In Fig.~2a, we show the real part of the electron self-energy at 70 K 
for a superconducting Bi$_{2}$Sr$_{2}$CaCu$_{2}$O$_{8+y}$ crystal with 
$T_{c}$ 
= 91 K.  The data are reproduced from Ref.~\cite{Johnson}. The 
electron self-energy  data are 
obtained from an ARPES spectrum, which is taken with an energy 
resolution of 
about 10 meV (Ref.~\cite{Johnson}). By analogy to the  Re$\Sigma$ data 
of Be(10$\bar{1}$0) obtained with the same energy resolution \cite{Shi}, we speculate that 
the  uncertainty of the Re$\Sigma$ data of BSCCO should be about $\pm$1
meV. It is apparent that the fine structures of 
electron-boson coupling also 
show up in  the electron self-energy of BSCCO. The 
energy of the highest peak is 66.0 meV, which is comparable with 
the energies (64-67 meV)
of the kink features in the band dispersion curves of 
several overdoped BSCCO crystals \cite{Gromko}. The solid line is a smooth curve obtained 
using the cubic spline interpolation with $\Lambda$ = -4.  Because the energy resolution for the BSCCO data is two times 
better than that for the LSCO data, it is reasonable to choose $\Lambda$ =
-4  and -2.5 to smooth the BSCCO 
and LSCO data, respectively. 

Taking the second derivative of the smoothed curve yields a 
$-$d$^{2}$(Re$\Sigma$)/d$\omega^{2}$ spectrum, which is shown in 
Fig.~2b. Here we have assigned $\omega$ = $E_{F}-E - 
\Delta_{D}$ with $\Delta_{D}$ = 7 meV. There are 6 pronounced peak features 
at 5.8 
meV, 21.4 meV, 38.3 meV, 44.4 meV, 60.0 meV, and 76.2 meV, which are
marked by vertical dashed lines.  It is remarkable that the peak
positions at 44.4 meV, 60.0 meV, and 76.2 meV for BSCCO match
precisely with the peak positions (marked by upward arrows) at 45 meV, 60 meV, and 76 meV, which are
consistently seen in the normal-state $-$d$^{2}$(Re$\Sigma$)/d$\omega^{2}$
spectra of underdoped superconducting LSCO crystals \cite{Zhou}. The peak at 38.3 
meV for BSCCO also lines up with the peak at 38.8 meV for La$_{1.97}$Sr$_{0.03}$CuO$_{4}$.
Such excellent agreement indicates that the structures at 38.3 meV, 44.4 meV, 60.0 meV, and 76.2 meV
are caused by strong coupling to the CuO$_{2}$-plane related phonon modes
and that the diagonal gap for BSCCO is 7$\pm$1 meV at 70 K. 

In order to further show that 
the structures in  $-$d$^{2}$(Re$\Sigma$)/d$\omega^{2}$ are 
indeed associated with strong electron-phonon coupling, we 
compare these structures with those in the phonon 
density of states. It is striking that the peak features in 
the $-$d$^{2}$(Re$\Sigma$)/d$\omega^{2}$ spectrum of  
Bi$_{2}$Sr$_{2}$CaCu$_{2}$O$_{8+y}$ match well with those in 
the phonon density of states obtained with neutron scattering.

In Fig.~3, we compare the $-$d$^{2}$(Re$\Sigma$)/d$\omega^{2}$ spectrum
with the $-$d$\bar{g}$/d$\omega$ tunneling spectrum (Ref.~\cite{Zhao06}).
Here $\bar{g}$ is the renormalized tunneling conductance~\cite{Zhao06}.
One can clearly see that all the peak structures in 
$-$d$^{2}$(Re$\Sigma$)/d$\omega^{2}$ of BSCCO line up precisely with 
those in the tunneling spectrum. Such quantitative agreement cannot be
caused by any artifacts and must indicate that the fine structures in 
both $-$d$^{2}$(Re$\Sigma$)/d$\omega^{2}$ and $-$d$\bar{g}$/d$\omega$ are
due to strong electron-phonon coupling.

\begin{figure}[htb]
    \includegraphics[height=6.2cm]{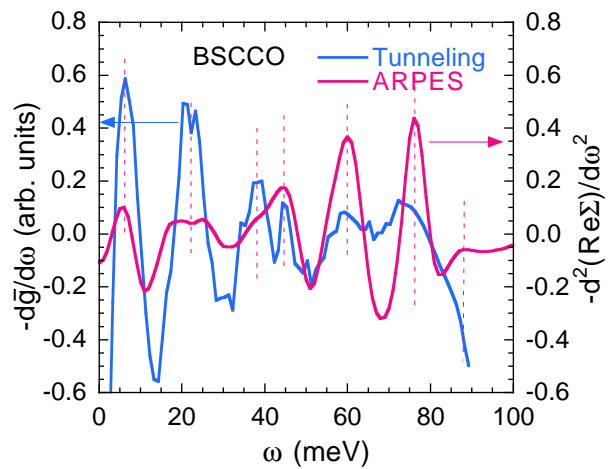}
	\caption[~] {The 
$-$d$^{2}$(Re$\Sigma$)/d$\omega^{2}$ spectrum and the $-$d$\bar{g}$/d$\omega$ tunneling 
spectrum of BSCCO crystal \cite{Zhao06}. The $-$d$\bar{g}$/d$\omega$ tunneling 
spectrum is reproduced from Ref.~\cite{Zhao06}. Vertical dashed lines mark the 
peak positions  in  the $-$d$^{2}$(Re$\Sigma$)/d$\omega^{2}$ spectrum. }
\end{figure}

The substantial non-zero diagonal gap 
observed in the nearly optimally doped BSCCO rules 
out $d$-wave gap symmetry and strongly 
supports an extended $s$-wave gap symmetry with eight line nodes. The extended $s$-wave gap symmetry is also 
consistent with many other 
independent experiments \cite{Zhao,Brandow}.  These include 
phase-sensitive experiments based on out-of-plane Josephson tunneling 
\cite{Li,Klemn,Sun}, single-particle tunneling spectroscopy 
\cite{Mag}, Raman 
spectroscopy of heavily overdoped cuprates \cite{Kend}, nonlinear 
Meissner effect \cite{Bha}. Moreover, the measurements of the physical 
properties that 
are related to low energy quasiparticle excitations 
\cite{Hardy,Jacobs,Lee,Chiao} are in quantitative agreement with
an extended $s$-wave gap symmetry with eight line nodes \cite{Zhao}.

Now we discuss the validity of the above analyses. The basic assumption of 
the analyses is that the normal-state
spectra display peak features at the characteristic phonon energies, and
that the superconducting-state spectra exhibit peak features at the same
energies shifted by the magnitude of the gap $\Delta_{D}$ along the diagonal
direction.  Our assumption for the superconducting state is not justified for a superconductor
with a strongly anisotropic superconducting gap if the coupling to the
large-momentum phonon modes dominates.  On the other hand, if the coupling to large-mementum
phonons are dominant, as in the case of conventional superconductors, the superconducting gap must be isotropic since 
exchange of phonons with 
large momenta effectively connects all points of the Fermi
surface.  In order to obtain a strongly anisotropic superconducting gap
within strong electron-phonon coupling, the coupling to small momentum
phonons must be dominant \cite{Abri}. This is indeed the case for the ionic-crystal-like cuprate
superconductors where the Coulomb screening is weak \cite{Abri}.  Then, the initial and
scattered electronic states connected by small momenta of phonons are close to each other so that the strong
electron-phonon coupling features in the
superconducting-state self-energy will be shifted by an angle-dependent gap. This
is especically true for the spectra along the diagonal and anti-nodal directions where
the superconducting density of states at $\Delta_{D}$ and $\Delta_{M}$
is singular in the case of the extended $s$-wave gap. This is because 
singular superconducting density of states at the diagonal gap makes the weight of the zero-momentum phonon 
scattering (connecting the initial diagonal to the scattered diagonal electronic states) dominate.

In summary, we have analyzed the data of electron self-energy along 
the diagonal direction  of a nonsuperconducting 
La$_{1.97}$Sr$_{0.03}$CuO$_{4}$ crystal and a superconducting 
Bi$_{2}$Sr$_{2}$CaCu$_{2}$O$_{8+y}$ crystal with 
$T_{c}$ 
= 91 K.  Our analyses clearly show that 
the diagonal superconducting gap at 70 K is
7$\pm$1 meV for the optimally doped BSCCO. The results rule 
out seemingly well accepted $d$-wave gap symmetry and strongly 
supports an extended $s$-wave gap symmetry with eight line nodes. The 
present work places an essential constraint on the microscopic 
pairing mechanism of high-temperature superconductivity.

 ~\\
~\\
$^{*}$Correspondence should be addressed to gzhao2@calstatela.edu

\end{document}